\documentclass[nonblindrev]{informs2}% no doublespacing for arxiv
\LRHSecondLine{}
\RRHSecondLine{}
\let\theARTICLETOPLEFT=\relax

\usepackage{endnotes}
\let\footnote=\endnote

\usepackage{enumerate}

\DeclareMathOperator\proj{proj}

\newcommand{\R}{\mathbb R}
\newcommand{\Z}{{\mathbb Z}}
\newcommand{\N}{{\mathbb N}}

\newenvironment{Proof}
  {{\sc Proof.}}
  {\hfill\Halmos}

\def\ve#1{\mathchoice{\mbox{\boldmath$\displaystyle\bf#1$}}
{\mbox{\boldmath$\textstyle\bf#1$}}
{\mbox{\boldmath$\scriptstyle\bf#1$}}
{\mbox{\boldmath$\scriptscriptstyle\bf#1$}}}

\renewcommand{\theenumi}{\roman{enumi}}

\newenvironment{inputlist}
  {\begin{enumerate}[\quad\rm({I}$_\bgroup 1\egroup$)]}
  {\end{enumerate}}
\newenvironment{outputlist}
  {\begin{enumerate}[\quad\rm({O}$_\bgroup 1\egroup$)]}
  {\end{enumerate}}

\usepackage{natbib}
 \bibpunct[, ]{(}{)}{,}{a}{}{,}%
 \def\bibfont{\small}%

\TheoremsNumberedThrough     % Preferred (Theorem 1, Lemma 1, Theorem 2)
\ECRepeatTheorems

\EquationsNumberedThrough    % Default: (1), (2), ...

\begin{document}

 \RUNAUTHOR{K\"oppe, Ryan, and Queyranne}

\RUNTITLE{Generating functions and IP games}

\TITLE{Rational Generating Functions and\\
  Integer Programming Games}

\ARTICLEAUTHORS{%
\AUTHOR{Matthias K\"oppe}
\AFF{University of California, Davis, Department of
  Mathematics, One Shields Avenue, Davis, CA 95616, USA\\
  \EMAIL{mkoeppe@math.ucdavis.edu}}%, \URL{}}

\AUTHOR{Christopher Thomas Ryan, Maurice Queyranne}
\AFF{University of British Columbia, Sauder School of Business,
2053 Main Mall, Vancouver, BC, Canada, V6T 1Z2 \\ \EMAIL{\{chris.ryan, maurice.queyranne\}@sauder.ubc.ca}}
} % end of the block

\ABSTRACT{%
We explore the computational complexity of computing pure Nash equilibria for a new class of strategic games called integer programming games with difference of piecewise linear convex payoffs. Integer programming games are games where players' action sets are integer points inside of polytopes. Using recent results from the study of short rational generating functions for encoding sets of integer points pioneered by Alexander Barvinok, we present efficient algorithms for enumerating all pure Nash equilibria, and other computations of interest, such as the pure price of anarchy, and pure threat point, when the dimension and number of ``convex" linear pieces in the payoff functions are fixed. Sequential games where a leader is followed by competing followers (a Stackelberg--Nash setting) are also considered.
}%


\makeatletter
\let\savemaketitle=\maketitle
\let\save@maketitle=\@maketitle
\maketitle

\section{Introduction}

In this paper we introduce a new class of strategic games that have appealing properties, and whose set of pure Nash equilibria can be described in a convenient encoding by rational generating functions using techniques pioneered by \cite{Barvinok1994} and expanded by \cite{Barvinok2003}. Generating functions techniques based on Barvinok's results have been applied to discrete optimization (see, for instance, \cite{deloera-hemmecke-koeppe-weismantel:mixedintpoly-fixeddim-fullpaper,DeLoera2007}), and various other areas of applied and pure mathematics (see \cite{DeLoera2005count} for a survey). To the authors' knowledge this is the first application of the theory of rational generating functions to the study of games.

This paper is motivated by open questions regarding the computational complexity of deciding the existence of pure Nash equilibria in strategic games. For a general reference on complexity issues in game theory see \cite{Papadimitriou2007}. As opposed to the case of mixed-strategy Nash equilibria which are guaranteed to exist for every game, the general problem of deciding if a strategic game has a pure Nash equilibrium is NP-hard (\cite{Gottlob2005}). In view of this difficulty, the problem has been explored under various restrictions on the actions and payoffs of the players; for instance, in graphical games (\cite{Alvarez2005,Gottlob2005}), congestion games (\cite{Dunkel2008,fabrikant-papadimitriou-talwar:2004:pure-nash}) and action graph games (\cite{Jiang2007}). This paper continues this tradition, by introducing a class of games that will be shown to have convenient algorithms to decide if instances have pure Nash equilibria, and if they exist, to compute them.

We consider \emph{integer programming games}, which are simultaneous games where the players' actions (pure strategies) are lattice points (i.e., integer points) inside polytopes described by systems of linear inequalities. Since the sets of actions are given implicitly by the description of polytopes, they may be of exponential size with respect to the input size. In our setting, each player's payoffs are given as \emph{difference of piecewise linear convex (DPLC) functions}. As an aside, optimization problems involving the difference of convex functions are a well-studied class of nonconvex programs (see for example \cite{Horst1999}).

The main  result of the paper is that the set of pure Nash equilibria of integer programming games with DPLC payoffs can be encoded as a short rational generating function in polynomial time when the number of players, dimensions of the polytopes that define the action sets and the number of ``convex" linear pieces in the payoffs are all fixed. Although these conditions are restrictive, note that each player may have an exponential number of actions. Indeed integer programming games with DPLC payoffs are a subset of a general class of games where deciding if a pure Nash equilibrium exists is $\Sigma_2^p$-complete with a fixed number of players and exponential-sized strategy spaces \cite{Alvarez2005}.

Besides questions of complexity, a short rational generating function encoding is a convenient data structure for answering other questions of interest regarding the structure of pure Nash equilibria and related concepts. For instance, several questions analogous to those explored in \cite{Conitzer2003} regarding \emph{mixed} strategy Nash equilibria can be answered efficiently in our setting for \emph{pure} Nash equilibria by using the rational generating function data structure.

We feel the main contributions of the paper are:
\begin{itemize}
\item Introducing the use of Barvinok short rational generating functions to the study of strategic games and demonstrating the power of encoding sets of pure equilibria as generating functions.
\item Presenting a tractable class of games, integer programming games with DPLC payoffs, for which pure strategy Nash equilibria and related quantities can be computed in polynomial time when certain dimensions are fixed.
\end{itemize}

Also of note are two ideas used in several places in this paper:
\begin{itemize}
\item In order to represent sets of equilibria, or other sets of interest, as rational generating functions we express the set as an overall feasible set in which unwanted elements, expressed as the union of projections of lattice point sets in polytopes, are removed. See for instance the proof of Theorem~\ref{theorem:extended-main-result} where the set of pure Nash equilibria is defined in this way. This is a general technique that is adapted to various settings in Sections~\ref{s:pure-nash}~to~\ref{s:stack-nash}.
\item Some results are easier to show when the actions for each player in the game are \emph{extended} to include a component that captures the payoff of that player.  This extension allows for descriptions of \emph{extended} strategy profiles and equilibria that are more amenable to generating function techniques and can readily be converted back to normal strategy profiles and equilibria. See for instance Definition~\ref{def:extended-game}.
\end{itemize}

The paper is organized into the following sections. Section \ref{s:integer-programming-games} introduces integer programming games and discusses an application of this framework to a model of competing firms producing identical indivisible goods. Section \ref{s:gen-fun} reviews the basic theory of Barvinok generating functions and major results that will be used in the remainder of the paper. Section \ref{s:pure-nash} discusses pure Nash equilibria and contains the main contributions of the paper -- demonstrating how generating functions can be used to encode sets of pure Nash equilibria. Section \ref{s:computations} details several other applications of generating function constructions to the computation of Pareto optima, the price of anarchy, and pure minmax values. Lastly, Section \ref{s:stack-nash} describes a sequential (Stackelberg--Nash) version of an integer programming game where a leader's actions affects the description of the polytopes defining the actions sets of a group of followers, who then play a simultaneous game.

\section{Integer Programming Games}\label{s:integer-programming-games}

We begin by introducing the following class of strategic games:

\begin{definition}[\textbf{Integer Programming Game}]
An \emph{integer programming game} with $n$ players is a noncooperative game where the $S_i$ of actions for each player $i$ is the set of lattice points inside a polytope; that is,
\begin{equation}\label{eq:strategySets}
S_i=P_i\cap\Z^{d_i}
\end{equation}
where $P_i = \{\ve x \in \R^{d_i}:M_i\ve x \le \ve b_i\}$ is a rational polytope.

Let $I = \{1,\dots,n\}$ denote the set of players. The set $S$ of action profiles $\ve s = (\ve s_1,\ldots,\ve s_n)$ is the Cartesian product of the $S_i$'s:
$$S=\prod_{i=1}^n S_i \subseteq \Z^d$$
where $d = d_1+\cdots+d_n$. The payoff functions are integer-valued of the form $u_i:S\rightarrow \Z$ for $i \in I$.
\end{definition}

As noted in the introduction, a distinguishing feature of this class of games is that the action sets $S_i$ are defined succinctly by linear systems $M_i\ve x \le \ve b_i$, even though $|S_i|$ may be exponential in size with respect to the input size of $M_i$ and $b_i$. We use rational generating functions to avoid explicitly enumerating each player's action set.

\begin{definition}[\textbf{DPLC payoffs}]
An integer-valued payoff function $u:S\rightarrow \Z$ of a game is a \emph{difference of piecewise-linear convex functions} or \emph{DPLC} function if it can be expressed as:
\begin{equation*}
u(\ve s) = \max_{k\in K} f_k(\ve s) - \max_{l \in L} g_l(\ve s)
\end{equation*}
where the $f_k$ and $g_l$ are affine functions with integer coefficients and where $K$ and $L$ are finite index sets. We refer to the $f_k$ as the ``convex" pieces of $u$ and the $g_l$ as the ``concave" pieces of $u$.
\end{definition}

We consider integer programming games where each player $i$ has a DPLC payoff function
\begin{equation}\label{eq:DPLC}
u_i(\ve s) = \max_{k\in K_i} f_{ik_i}(\ve s) - \max_{l \in L_i} g_{il_i}(\ve s)
\end{equation}
again where the $f_{ik}$ and $g_{il}$ are given affine functions with integer coefficients and the sets $K_i$ and $L_i$ are given finite index sets.

A first comment regarding this class of games is that they can express any finite game given in normal form. A normal form game is defined by action sets $A_1,\ldots,A_n$ and payoffs $\pi_i(\ve a)$ for $\ve a \in A_1\times\cdots\times A_n$ and $i \in I$. We refer to the set $A = \prod_{i=1}^n A_i$ as the set of action profiles. A normal form game is finite is $A$ is finite. We use the alternate notation $A_i$ and $\pi_i$, as opposed to $S_i$ and $u_i$, for normal form games to draw a contrast between the fact that in normal form game the actions sets and payoffs for each action profile are given explicitly as part of the input, whereas in integer programming games the action sets $S_i$ and payoffs $u_i$ are given implicitly. This contrast between implicit and explicit representations is important when one considers the computational complexity of finding pure Nash equilibria (see \cite{Alvarez2005} for a technical discussion).

\begin{proposition}\label{prop:expressibility}
Every finite normal form game is equivalent to an integer programming game with DPLC payoffs.
\end{proposition}
\begin{Proof}
Let finite normal form game $G$ be defined by action sets $A_1,\ldots,A_n$ and payoffs $\pi_i(\ve a)$ for $\ve a \in A$.  Let $d_i$ equal the number of elements in the action set $A_i$. Let vector $\ve x_i =(x_{i,1},\ldots,x_{i,d_i})$ denotes a mixed strategy profile for player $i$. Mixed strategy profiles $\ve x$ lie in the unit simplex:
$$P_i=\{\ve x_i=(x_{i,1},\ldots,x_{i,d_i}) \in \R^{d_i}:\ve x_i \ge 0, ~\sum_{j=1}^{d_i} x_{i,j}=1\}.$$
This is a polytope, and the lattice points inside $P_i$ are exactly its vertices, which correspond to pure strategies (actions) for player $i$, namely, an action $a_i$ in $A_i$ is represented by its characteristic vector $\chi_i(a_i)$. Thus, we represent the set of pure strategies for player $i$ as $S_i = P_i \cap \Z^{d_i}$. As above let $S = S_1 \times \dots \times S_n$. This conforms to the framework of integer programming games.

As for the payoffs, our goal is to convert the explicitly given payoffs $\pi_i$ over the set $A$ of action profiles into equivalent DPLC payoff functions $u_i$ over $S$. Let $A_i^+=\{\ve a\in A:\pi_i(\ve a)>0\}$ and $A_i^-=\{\ve a\in A:\pi_i(\ve a)<0\}$. For every $\ve x = (\chi_1(a_1),\dots,\chi_n(a_n)) \in S$, define the DPLC payoff
$$u_i(\ve x) = \max_{a_i \in A_i^+} \pi_i(\ve a) \left(\sum_{i=1}^n x_{i,a_i} - n + 1\right) - \max_{a_i \in A_i^-}\pi_i(\ve a) \left( n-\sum_{i=1}^n x_{i,a_i} - 1\right).$$
Note that for every $\ve a \in A$ we have $u_i(\chi_1(a_1),\dots,\chi_n(a_n))=\pi_i(\ve a)$.
Thus the normal form game $G$ is equivalent to the integer programming game with pure strategy sets $S_i$ and DPLC payoffs $u_i$.
\end{Proof}

Note that this representation has the same input size as the normal form game itself. Further computational complexity consequences of this proposition are discussed in Remark~\ref{remark:expressibility}.

Another useful result demonstrates the power of modeling payoffs as DPLC functions. \citet[Theorem 4.2]{Zalgaller2000} shows that every continuous piecewise linear function can be represented as the difference of two piecewise linear convex functions. Thus, a DPLC function can be used to describe any continuous piecewise linear payoff function, which in turn can be used to approximate an arbitrary continuous payoff function.

\begin{example}[\textbf{Cournot game with indivisible goods and fixed production costs}]\label{example}~\\
A set of $n$ manufacturers (the players) produce indivisible goods which are then sold in a market. Player $i$ chooses an integer production level $s_{ij}$ for each of its indivisible products $j \in \{1,\dots,d_i\}$ subject to resource constraints $M_i \ve s_i \le \ve b_i$ where $\ve s_i = (s_{i1},\ldots,s_{id_i})$. Thus, player $i$'s action set $S_i$ is the set of integer points in a polytope in $\R^{d_i}$. The payoff to each player consists of revenues and production costs. Under usual assumptions, the revenue to each player $i$ is a concave function of the action profile $\ve s$, which can be expressed as a piecewise linear concave function $\min_{l \in R_i} r_{il}(\ve s)$. For each player $i$ and product $j$ there may be a fixed production cost $F_{ij}$. The variable production cost is a convex function of the production levels $\ve s_i$ expressed as a piecewise linear convex function $\max_{l \in C_i} c_{il} (\ve s_i)$. The payoff to player $i$ is thus
\begin{displaymath}
u_i(\ve s) = \sum_{j=1}^{d_i} \max\{-F_{ij}, -F_{ij}s_{ij}\} - \left(\max_{l \in R_i} (-r_{il}(\ve s)) + \max_{l \in C_i} c_{il} (\ve s_i) \right)
\end{displaymath}
which can be expressed as a DPLC function. As will be discussed further in Remark~\ref{remark:example}, this is precisely the structure that is analyzed in this paper.
\end{example}

\section{Introduction to rational generating functions}\label{s:gen-fun}

We give a very brief overview of the theory of short rational generating
functions used in this paper. \cite{Barvinok1994}
introduced an algorithm for determining the exact
number of lattice points in a rational polytope $P= \{\, \ve x \in \R^n : A
\ve x \leq \ve b \,\}\,$, that runs in polynomial time for every fixed
dimension~$n$.   This algorithmic breakthrough provided a strengthening
of the famous algorithm of \cite{Lenstra83}, which allows to
\emph{decide} whether $P$ contains a lattice point in polynomial time for
every fixed dimension.

Barvinok's method works as follows.  Consider the \emph{generating
  function} of the lattice point set~$P\cap\Z^n$, which is defined as the
multivariate Laurent polynomial
\begin{equation}\label{eq:long-generating-function}
  g(P;\ve\xi) = \sum_{\ve x\in P\cap\Z^n} \ve\xi^{\ve x}
  = \sum_{\ve x\in P\cap\Z^n} \xi_1^{x_1}\cdots \xi_n^{x_n} \in \Z[\xi_1^{\pm1},\dots,\xi_n^{\pm1}].
\end{equation}
This Laurent polynomial, when encoded as a list of monomials, has exponential
encoding size.  Barvinok's algorithm computes a different representation of
the function $g(P;\ve\xi)$ as a sum of basic rational functions in the form
\begin{equation}\label{eq:barvinok-formula}
  g(P;\ve\xi) = \sum_{i \in I} \gamma_i
  \frac{\ve \xi^{\ve c_i}}{(1-\ve \xi^{\ve d_{i1}})(1-\ve \xi^{\ve d_{i2}})\dots
    (1-\ve \xi^{\ve d_{in}})},
\end{equation}
where $I$ is a polynomial-size index set and all data are integer.  This
algorithm runs in polynomial time whenever the dimension~$n$ is a fixed
constant.  A formula of the type~\eqref{eq:barvinok-formula} is called a
\emph{short rational generating function}.

Note that each of the basic rational functions has poles (the
point~$\ve\xi=\ve 1$ in particular is a pole of all the basic rational
functions), but after summing up only removable singularities remain.
Obtaining the exact number of lattice points of~$P$ is easy
in~\eqref{eq:long-generating-function}, since clearly $|P\cap\Z^n| =
g(P;\ve1)$.  Since~\eqref{eq:barvinok-formula} is a formula for the same
function (except for removable singularities), we also have $|P\cap\Z^n| =
\lim_{\ve\xi\to\ve1} g(P;\ve\xi)$, which can be evaluated in polynomial time by performing
a residue calculation with each basic rational function in the
sum~\eqref{eq:barvinok-formula}.
An important point to note is that this evaluation is possible with arbitrary
rational generating functions that correspond to finite lattice point sets.
In other words, if we can compute in polynomial time a rational generating function of
a finite lattice point set~$S\subseteq\Z^n$, we can also compute in polynomial time its
cardinality $|S|$.  Therefore, we can also decide in polynomial time whether $S \not =\emptyset$.

A first, trivial observation that allows to combine rational generating functions is
the following.  Let $X\subseteq\Z^n$ and $Y\in\Z^k$ be lattice point sets
given by their short rational generating functions~$g(X; \ve\xi)$ and
$g(Y;\ve\eta)$.  Then the direct product (Cartesian product) $X\times Y$ also
has a short rational generating function that is simply the product of the
rational functions:
\begin{displaymath}
  g(X\times Y; \ve\xi,\ve\eta) = g(X;\ve\xi) \times g(Y;\ve\eta).
\end{displaymath}

\cite{Barvinok2003} developed powerful algorithms to obtain
short rational generating functions of more general lattice point sets.
The first of these algorithms concerns constant-length Boolean combinations of
finite lattice
point sets that are already given by rational generating functions.
\begin{theorem}[Boolean Operations Theorem] \label{theorem:boolean-operations}
  \emph{(Corollary 3.7 in
  \cite{Barvinok2003})} Let $m$ and $\ell$ be fixed integers,
  and let $\phi\colon \{0,1\}^m\to \{0,1\}$ be any Boolean function
  such that $\phi(\ve0) = 0$.
  %% (This restriction is there so that \phi(S_1,\dots,S_p) will be
  %% contained in the union of S_1,\dots,S_p, thus finite. --mkoeppe)
  Then there exists a constant $s = s(\ell, m)$ and
  a polynomial-time algorithm for the following problem.
  Given as \emph{input}, in binary encoding,
  \begin{inputlist}
  \item the dimension~$n$ and
  \item rational generating functions
    $$ g(S_p; \ve \xi) =  \sum_{i \in I_p} \gamma_{pi} \frac { \ve \xi^{\ve c_{pi}} } {  (1-\ve
      \xi^{\ve d_{pi1}}) \dots (1-\ve \xi^{\ve d_{pis}}) },$$ of $m$ finite
    sets~$S_p\subseteq\Z^n$, represented by the rational numbers $\gamma_{pi}$,
    integer vectors $\ve c_{pi}$ and $\ve d_{pij}$ for $p=1,\dots,m$, $i\in I_p$,
    $j=1,\dots,\ell_{mp}$ such that the numbers $\ell_{mp}$ of terms in the
    denominators are at most~$\ell$,
  \end{inputlist}
  \emph{output}, in binary encoding,
  \begin{outputlist}
  \item rational numbers $\gamma_i$, integer vectors $\ve c_i$, $\ve d_{ij}$ for
    $i\in I$, $j=1,\dots,s_i$, where $s_i \leq s$, such that
    $$ g(S; \ve \xi) =  \sum_{i \in I} \gamma_i \frac { \ve \xi^{\ve c_i} } {  (1-\ve
      \xi^{\ve d_{i1}}) \dots (1-\ve \xi^{\ve d_{is_i}}) }$$ is a rational generating
    function of the finite set~$S$ that is the Boolean combination of
    $S_1,\dots,S_p$ corresponding to the function~$\phi$.
  \end{outputlist}
\end{theorem}
Note that the restriction~$\phi(\ve0) = 0$ ensures that the set~$S$ will be finite.
The essential part of the construction of Theorem~\ref{theorem:boolean-operations} is the
implementation of set intersections, which are based on the
\emph{Hadamard product} \citep[Definition 3.2]{Barvinok2003}, which is the
bilinear extension of the operation defined on monomials as
\begin{displaymath}
  \alpha\ve x^{\ve\xi} * \alpha' \ve x^{\ve\xi'}
  = \begin{cases}
    \alpha\alpha' \ve x^{\ve\xi} & \text{if $\ve\xi=\ve\xi'$,} \\
    0 & \text{otherwise.}
  \end{cases}
\end{displaymath}
With this definition, clearly
\begin{displaymath}
  g(S_1\cap S_2) = g(S_1;\ve\xi) * g(S_2;\ve\xi).
\end{displaymath}

Another powerful method to define lattice point sets is by integer
projections.
Let~$S\subseteq\Z^n$ be a finite lattice point set, given by its
rational generating function~$g(S;\ve\xi)$.  Let $\psi\colon \Z^n\to\Z^k$ be a
linear function and denote by~$T=\psi(S)$ the image (projection) of~$S$.
If the map~$\psi$ is
one-to-one (injective) from~$S$, then
the generating function~$g(T; \ve\eta)$ of the
projection~$T$ can be computed by making a \emph{monomial substitution} in
$g(S;\ve\xi)$; see \citet[Theorem~2.6]{Barvinok2003}. This fact is used in the
proof of Corollary~\ref{theorem:pure-nash-ratgenfun-encoding-concave}.

When $S$ is the set of lattice points in a polytope~$P$,
the integer projection method of \cite{Barvinok2003} can be
employed to construct a rational generating function of
the projection~$T$.

\begin{theorem}[Projection Theorem] \label{theorem:projection}
\emph{(Theorem 1.7 in \cite{Barvinok2003})} Let the dimension $n$ be a fixed constant.
Then there exists a constant~$s = s(n)$ and a polynomial-time algorithm for the
following problem.
Given as \emph{input}, in binary encoding,
\begin{inputlist}
\item an inequality description of a rational polytope~$P \subset\R^n$;
\item a positive integer~$k$; and
\item a linear map $\psi\colon\Z^n\to\Z^k$ given by an integral matrix;
\end{inputlist}
\emph{output}, in binary encoding,
\begin{outputlist}
\item rational numbers $\gamma_i$, integer vectors $\ve c_i$, $\ve d_{ij}$ for
  $i\in I$, $j=1,\dots,s_i$, where $s_i \leq s$, such that
$$ g(T; \ve \xi) =  \sum_{i \in I} \gamma_i \frac { \ve \xi^{\ve c_i} } {  (1-\ve
  \xi^{\ve d_{i1}}) \dots (1-\ve \xi^{\ve d_{is_i}}) }$$ is a rational generating
function of the set $T = \psi(P \cap \Z^n)$.
\end{outputlist}
\end{theorem}

Once a rational generating function of a set~$S$ has been computed, various
pieces of information can be extracted from it.  We have already mentioned that it is
possible to compute the cardinality of~$S$
In addition to that, we can \emph{explicitly enumerate} all elements of~$S$.  Since the
cardinality of~$S$ can be exponential in the encoding length of the input, we use
output-sensitive complexity analysis, i.e., to measure the
complexity of the enumeration algorithm in terms of both the input and the
output.  The strongest notion of an output-sensitive polynomial-time
enumeration algorithm is that of a \emph{polynomial-space polynomial-delay
  enumeration algorithm}.  Such an algorithm only uses space that is
polynomial in the encoding length of the input data.  In addition, the time
spent between outputting two items, and before outputting the first item and
after outputting the last item, is bounded by a polynomial in the encoding
length of the input data.  The following result is a version of Theorem~7 of
\cite{DeLoera2007}.

\begin{theorem}[{\bf Enumeration Theorem}] \label{theorem:output-sensitive-enumeration}
  Let the dimension $k$ and the maximum number~$\ell$ of binomials in the
  denominator be fixed.  Then there exists a polynomial-space polynomial-delay
  enumeration algorithm for the following enumeration problem.
  Given as \emph{input}, in binary encoding,
  \begin{inputlist}
  \item a number $M\in\Z_+$;
  \item rational numbers $\gamma_i$, integer vectors
    $\ve c_i$, $\ve d_{ij}$ for $i\in I$, $j=1,\dots,\ell_i$, where $\ell_i
    \leq \ell$ such that
    $$
    \sum_{i \in I} \gamma_i \frac{\ve s_0^{\ve c_i}}{(1-\ve s_0^{\ve d_{i1}})(1-\ve
      z^{\ve d_{i2}})\dots (1-\ve s_0^{\ve d_{is}})}
    $$
    is a %positively weighted
    rational generating function of a set $V\subseteq\Z^{k}$ of lattice
    points with $V\subseteq[-M,M]^{k}$;
  \item an integer~$p$ with $1\leq p \leq k$
  \end{inputlist}
  \emph{output}, in binary encoding,
  \begin{outputlist}
  \item all points in the projection of $V$ onto the last $p$~components,
    \[
    W = \{\, \ve w\in\Z^p: \exists \ve t\in\Z^{k-p} \text{ such that } (\ve
    t,\ve w)\in V \,\},
    \]
    in lexicographic order.
  \end{outputlist}
\end{theorem}

In addition, binary search can be used to optimize a linear function over a lattice point set encoded
as a rational generating function.

\begin{theorem}[{\bf Linear Optimization Theorem}] \label{theorem:linear-optimization-binary-search}
  Let the dimension $k$ and the maximum number~$\ell$ of binomials in the
  denominator be fixed.  Then there exists a polynomial-time algorithm
  for the following problem.
  Given as \emph{input}, in binary encoding,
  \begin{description}
  \item[~~~\emph{($\text{I}_1$)}] and \emph{($\text{I}_2$)} as in Theorem \ref{theorem:output-sensitive-enumeration}
  \item[~~~\emph{($\text{I}_3$)}] a vector $\ve f\in\Z^k$,
  \end{description}
  \emph{output}, in binary encoding,
  \begin{outputlist}
  \item an optimal solution~$\ve v^*\in\Z^k$ of the optimization problem
    $\max \{\, \langle \ve f, \ve v\rangle : \ve v\in V \,\}$.
  \end{outputlist}
\end{theorem}

We will use all the above results in the following constructions.

\section{Calculating Pure Nash Equilibria in Integer Programming Games}\label{s:pure-nash}

Consider an integer programming game with DPLC payoffs as defined in Section \ref{s:integer-programming-games}. Our goal is to encode the Nash equilibria of such a game as a short rational generating function. The most general setting we provide an efficient algorithm for such an encoding is when the number of players and the dimension of their action sets are fixed and each player's DPLC payoff function has the form:
\begin{equation}\label{eq:DPLC}
u_i(\ve s) = \max_{k\in K_i} f_{ik}(\ve s) - \max_{l \in L_i} g_{il}(\ve s)
\end{equation}
where we now assume that the size of $K_i$ is a fixed. Since $S$ is bounded we assume without loss of generality that $u_i(\ve s) \ge 0$ for all $i \in I$ and $\ve s \in S$. The analysis proceeds with two fundamental insights. First, when there are a fixed number of ``convex'' pieces, i.e.,  when $|K_i|$ is fixed, each player's payoff is piecewise linear concave within the region where an $f_{ik}$'s remains maximum. The second insight is that when payoffs are piecewise linear concave the hypograph of the payoff function is then a polyhedral set, encodable as a short rational generating function.

First, a simple result towards partitioning the action profile space into regions according to the values of the linear pieces of the payoffs. We assume that $K_i$ is a totally ordered set.

\begin{lemma} For each player $i$, the set of all action profiles can be expressed as a disjoint union
$$S = \biguplus_{k \in K_i} S_{ik}$$
where
$$S_{ik} = \left\{\ve s \in S: \begin{array}{ll} f_{ik}(\ve s) \ge f_{ij}(\ve s), & ~j > k \\
												                                        f_{ik}(\ve s) > f_{ij}(\ve s), & ~j <k \end{array}
\right\}.$$
\end{lemma}
\begin{Proof}
We first show that $S = \bigcup_{k \in K_i} S_{ik}$ and later establish it is a disjoint union. Clearly $ \bigcup_{k \in K_i} S_{ik} \subseteq S$. It remains to show the reverse inclusion. Let $\ve s \in S$ and define $J (\ve s)  = \{j \in K_i: f_{ik}(\ve s) = \max_{k \in K_i} f_{ik}(\ve s)\}$ and $j(\ve s) = \min J(\ve s)$, then $\ve s \in S_{ij}(\ve s)$.

To show that the union is in fact disjoint suppose by contradiction that there exists an $\ve s$ in $S_{ik}\cap  S_{ik'}$ where $k > k'$. If $k>k'$ then since $\ve s \in S_{ik}$ this implies $f_{ik}(\ve s) \ge f_{ik'}(\ve s)$. However, since $\ve s \in S_{ik'}$ this yields that $f_{ik'}(\ve s) > f_{ik}(\ve s)$, a contradiction. The result follows.
\end{Proof}

Note that we could equivalently write $S_{ik}$ as follows,
$$S_{ik} = \left\{\ve s \in S: \begin{array}{ll} f_{ik}(\ve s) \ge f_{ij}(\ve s), & ~j > k \\
												                                        f_{ik}(\ve s) \ge f_{ij}(\ve s) + 1, & ~j <k \end{array}\right\}.$$
We can do this because the action profiles in $S$ are integer vectors and the data defining the functions $f_{ij}$ are integral. The same holds true for all the strict inequalities in this paper, and thus the alternative descriptions using a strict inequality or a weak inequality with one unit added to the larger side of the inequality are used interchangeably throughout.

The next step is to refine the partition of $S$ to account for all players simultaneously. To do so we introduce the following convenient notation. Let $\ve K = \prod_{i=1}^n K_i$ be the set of all vectors $\ve k = (k_1,\ldots,k_n)$ of indices where $k_i \in K_i$ for $i \in I$. Using this notation, denote by $S_{\ve k}$ the intersection $S_{1k_1}\cap\cdots\cap S_{nk_n}$.
Employing this notation we state the following simple corollary of the previous lemma:

\begin{corollary}\label{cor:partition}
The set of action profiles can be partitioned as follows:
$$S = \biguplus_{\ve k \in \ve K} S_{\ve k} $$
\end{corollary}

Thus each action profile $\ve s$ lies in a unique subset $S_{\ve k}$ of $S$ where it is known that the payoff for each player $i$ is
$$u_i(\ve s) = f_{ik_i}(\ve s) + \min_{l \in L_i} g_{il}(\ve s)$$
and hence a piecewise linear concave function of $\ve s$. To take advantage of this concave structure in the payoff we propose an extension of the game.

\begin{definition} [\textbf{Extended Game}] \label{def:extended-game}
Given an integer programming game with DPLC payoffs let $\ve {\hat s} = (\ve s;\ve y) = (\ve s_1, \ldots, \ve s_n; y_1,\ldots, y_n)$ denote an \emph{extended action profile} which includes variables $y_i \in \Z$ which keep track of payoff values. The  \emph{set $\hat S$ of extended action profiles} is the set
\begin{equation}\label{eq:ShatUnion}
\hat S = \biguplus_{\ve k \in \ve K} \hat S_{\ve k}
\end{equation}
where
$$\hat S_{\ve k} = \{\ve {\hat s}=(\ve s, \ve y): \ve s \in  S_{\ve k}  \text{ and } 0 \le y_i \le f_{ik}(\ve s) - g_{il}(\ve s) \text{ for all } l \in L_i \text{ and } i \in I \}.$$
By Corollary~\ref{cor:partition} it is easy to see that \eqref{eq:ShatUnion} is a disjoint union.

We define the \emph{extended utility function} $\hat{u}_i$ for each player $i$ as
$$\hat{u}_i(\hat {\ve s}_i) = y_i$$
which is a linear function on $\hat S$.
We call the tuple $\hat G = (\hat S, \hat u_1,\ldots, \hat u_n)$ the \emph{extended game} of the original game $G = (S,u_1,\ldots,u_n)$.
\end{definition}

At this point there are three important observations regarding the definition of extended games. Note that since $S$ is a bounded by a polytope and the constraints $0 \le y_i \le f_{ik}(\ve s) - g_{il}(\ve s) \text{ for all } l \in L_i \text{ and } i \in I$ bound the values of $y_i$ it follows that each $\hat S_{\ve k}$, and thus $\hat S$ is bounded. This fact will be important when encoding $\hat S$ by a rational generating function and applying the Linear Optimization Theorem, which we have occasion to do in what follows. Note also that $\hat u_i(\hat {\ve s_i})$ is a linear function over $\hat S$ and thus more amenable to encoding by generating functions than the piecewise linear payoffs of the original game.

We also remark that the extended game $\hat G$ is not a simultaneous-move game since the players' choices are not independent and some choices of $\hat {\ve s}$ may lead to infeasibility. Similar issues are explored in \cite{Bhattacharjee2000}; however, we do not treat the extended game $\hat{G}$ as a game unto itself, but simply as a mathematical construct to study the equilibria of the original game $G$, and we thus ignore these considerations.

A key step is to establish a correspondence between the pure Nash equilibria to those of the original game. As will be seen, this correspondence relies on the fact that the descriptions of the action profile sets involve disjoint unions.

In analyzing games we often consider deviations from the vector $\ve k = (k_1,\ldots, k_n)$ in which the $i$-th component $k_i$ is replaced with $k'_i \in K_i$. For convenience, let $\ve k_{-i}$ denote the vector of the remaining (unchanged) indexes; therefore,
$$(\ve k_{-i},k'_i) = (k_1,\ldots,k_{i-1},k'_i,k_{i+1},\ldots,k_n).$$
We follow a similar convention in terms of action profiles. Namely, given an action profile $\ve s = (\ve s_1,\ldots,\ve s_n)$, if player $i$ deviates to action $\ve s'_i$ then we let $\ve s_{-i}$ denote the (unchanged) actions of the remaining players, so
$$(\ve s_{-i},\ve s'_i) = (\ve s_1,\ldots,\ve s_{i-1},\ve s'_i,\ve s_{i+1},\ldots,\ve s_n).$$

The equilibrium concept used in the extended game is defined as follows:

\begin{definition}[\textbf{Extended pure Nash equilibrium}]
Let $\hat G = (\hat S, \hat u_1,\ldots, u_n)$
be an extended game.
An \emph{extended pure Nash equilibrium} $\hat {\ve s} =(\ve s;\ve y)$ is an
extended action profile where if $\hat{\ve s} \in \hat S_{\ve k}$ then there
does not exist $\hat {\ve s}_i^\prime = (\ve s_i^\prime, y_i^\prime)$ such that $(\hat {\ve s}_{-i}, \hat {\ve s_i}^\prime) \in \hat S_{(\ve k_{-i}, k_i^\prime)}$ and $\hat u_i(\hat{\ve s}_{-i},\hat{\ve s_i}^\prime) > \hat u_i(\ve s).$
\end{definition}

\begin{lemma}\label{lemma:bijection}
Consider the game $G = (S,u_1,\ldots, u_n)$ and its extended game $\hat{G} = (\hat{S}, \hat{u}_1, \ldots, \hat{u}_n)$ as defined above.
\begin{enumerate}[\rm(i)]
\item An extended pure Nash equilibrium of $\hat{G}$ must be of the form
\begin{equation}\label{eq:form} \hat{\ve s} = (\ve s;\ve u(s))=(\ve s_1,\ldots,\ve s_n;u_1(\ve s),\ldots,u_n(\ve s)). \end{equation}
\item There is a bijection between the set $N$ of pure Nash equilibria of the original game and the set $\hat N$ of extended pure Nash equilibria of the extended game.
\end{enumerate}
\end{lemma}

\begin{Proof}
(i) Let $\hat{\ve s} = (\ve s; \ve y) \in \hat{S}$. By the disjoint union \eqref{eq:ShatUnion} there exists a unique $\ve k$ such that $\hat {\ve s} \in \hat S_{\ve k}$. It follows that $\ve s \in S_{\ve k}$. For all $i\in N$ we have $\ve s \in S_{ik_i}$ and $y_i  \le  f_{ik_i}(\ve s) - \max_{l \in L_i} g_{il}(\ve s)  =  u_i(s).$
Thus, $\hat u_i(\hat {\ve s}) = y_i \le u_i(\ve s) = \hat u_i(\ve s_1,u_1(\ve s);\ldots;\ve s_n,u_n(\ve s))$.
Hence, whenever $y_i < u_i(\ve s)$ it is profitable for player $i$ to deviate to the extended action $(\ve s_i;u_i(\ve s))$.
Therefore, an extended pure Nash equilibrium must have the form \eqref{eq:form}.
(ii) Consider the mapping $\varphi : N \longrightarrow \hat{N}$ defined by $\ve s \longmapsto (\ve s; u(\ve s))$. We claim $\varphi$ is a well-defined bijection.
First we show that $\varphi$ is well-defined, that is $\varphi(\ve s) \in \hat N$ for all $\ve s \in N$. Clearly $\varphi(\ve s)$ is in $\hat{S}$. Let $\ve s \in S_{\ve k}$ for some $\ve k$. Now consider a feasible deviating action for player $i$, $\hat {\ve s}_i^\prime = (\ve s_i^\prime, y_i^\prime)$, where $(\hat{ \ve s}_{-i}, \hat{\ve s}_i^\prime) \in \hat S_{(\ve k_{-i}, k_i^\prime)}$ for some $k_i^\prime \in K_i$. In other words, player $i$ deviates from choosing $\ve s_i \in S_{ik_i}$ to choosing $\ve s_i^\prime \in S_{ik_i^\prime}$ and changing $y_i$ to $y_i^\prime$. We have
$$\hat u_i(\hat{\ve s}_{-i},\hat{\ve s}_i^\prime) = y_i^\prime \le u_i(\ve s_{-i},\ve s_i^\prime) \le u_i(\ve s) = \hat u_i(\hat{\ve s})$$
where the second inequality holds since $\ve s$ is a Nash equilibrium for the game $G$ and the final equality follows from (i). It follows that $\hat{\ve s}$ is an extended Nash equilibrium for the game $\hat{G}$ and mapping $\varphi$ is well-defined.

It is clear that $\varphi$ is injective. As for surjectivity, by part (i) if follows that every Nash equilibrium in $\hat{N}$ has the form $(\ve s;u(\ve s))$ for some $\ve s \in S$. It just remains to show that all such equilibria arise with $\ve s \in N$. Suppose the contrary, that is $\hat{\ve s} = (\ve s;u(\ve s)) \in \hat{N}$ but $\ve s \not \in N$. Then there must be a profitable deviation $\ve s_i^\prime \in S_i$ for some player $i$ in the original game $G$;
$u_i(\ve s_{-i}, \ve s_i^\prime) > u_i(\ve s)$. This implies that there is a profitable deviation $\hat{\ve s}_i^\prime = (\ve s_i^\prime; u_i(\ve s_{-i}, \ve s_i^\prime))$ in the extended game since
$$\hat{u}_i(\hat{\ve s})= u_i(\ve s) < u_i(\ve s_{-i}, \ve s_i^\prime) = \hat{u}_i(\hat{\ve s}_{-i},\hat{\ve s}_i^\prime).$$
\end{Proof}

With this bijection, we can now state the main result of the paper:

\begin{theorem}\label{theorem:extended-main-result}
Consider an integer programming game with DPLC payoffs given by the following
input in binary encoding:
\begin{inputlist}
\item the number $n$ of players, and a bound $B \in \N$;
\item for each $i \in I = \{1,\dots,n\}$, the dimension $d_i$ and an inequality description $(M_i,\ve b_i)$ of a rational polytope $P_i = \{\ve x \in \R^{d_i}: M_i\ve x \le \ve b_i\}\subseteq [-B,B]^{d_i}$;
\item for each $i \in I$, nonnegative integers $|K_i|$ and $|L_i|$, and for all integers $k$, $l$ such that $1\le k \le |K_i|$ and $1\le j \le |L_i|$, integer vectors $\ve \alpha_{ik} \in \Z^{d}$, $\ve \gamma_{il} \in \Z^{d}$ (where $d = d_1 + \cdots + d_n$) and integers $\beta_{ik}$, $\delta_{ik}$   defining the affine functions $f_{ik}:S \rightarrow \Z$ and $g_{il}:S\rightarrow \Z$ by $f_{ik}(\ve s) = \ve \alpha_{ik}\cdot\ve s + \beta_{ik}$ and $g_{il}(\ve s) = \ve \gamma_{il}\cdot\ve s + \delta_{il}$ for all $\ve s\in S = \prod_{i=1}^n (P_i \cap \Z^{d_i})$.
\end{inputlist}
The set $\hat N$ of extended pure Nash equilibria of the extended game $\hat G = (\hat S, \hat u_1, \ldots, \hat u_n)$ has a short rational generating function encoding, which can be computed in polynomial time when the total dimension~$d$ and the sizes $|K_i|$ are fixed for all $i \in I$.
\end{theorem}
\begin{Proof}
We express the set of extended Nash equilibria as follows:
$$\hat N = \hat S \setminus \bigcup_{i=1}^n D_i$$
where
\begin{equation}\label{eq:deviation-set}
D_i = \biguplus_{\ve k\in \ve K}~\bigcup_{k_i^\prime \in K_i} \proj_{\hat {\ve s}}\left\{(\hat {\ve s}, \hat {\ve s}_i^\prime):
\hat {\ve s} \in \hat S_{\ve k},~ (\hat {\ve s}_{-i}, \hat {\ve s}_i^\prime) \in  \hat S_{(\ve k_{-i},k_i^\prime)},~ \hat u_i(\hat {\ve s}_{-i}, \hat {\ve s}_i^\prime) \ge \hat u_i(\hat {\ve s}) + 1
\right\}
\end{equation}
Note that some of the projected sets may be empty.

The set $D_i$ is the set of action profiles where player $i$ has a profitable deviation. The description of $D_i$ in \eqref{eq:deviation-set} is a union over profitable deviations from one set in the partition of $\hat S$ to another.  This description of $\hat N$ is easy to verify using the definition of extended pure Nash equilibria.

We now establish that $\hat N$ can be encoded as a short rational generating function. First we claim $\hat S$ admits such an encoding.  Consider the description of $\hat S$ given in \eqref{eq:ShatUnion}. The sets $F_{ik}$ are sets of lattice points inside rational polytopes and thus encodable as short rational generating functions. This in turn implies that $\hat S_{\ve k}$ admits such an encoding since it in turn is the set of lattice points inside a polytope. By the Boolean Operations Theorem, it follows that $\hat S$ can be encoded as a short rational generating functions in polynomial time, since there is a constant number of sets $\hat S_{\ve k}$ under the assumption that the sets $I$ and $K_i$ are of fixed size.

Note in addition that the sets to be projected in $\eqref{eq:deviation-set}$
are again sets of lattice points inside of rational polytopes, by observing
that the extended payoffs functions are linear.  By the Projection Theorem it
follows that each set in the union can be encoded as a short rational
generating function. Using again the Boolean Operations Theorem we conclude
that each $D_i$, and thus $\hat N$, admit short rational generating function
encodings which can be computed in polynomial time when the sizes of the sets
$K_i$ are fixed for all $i \in I$.  We remark that the outer union
in~\eqref{eq:deviation-set} (indexed by $\ve k\in \ve K$) is a disjoint union;
thus its rational generating function can be computed by adding the rational
generating functions of the parts, rather than using the construction of the
Boolean Operations Theorem.
\end{Proof}

\begin{corollary} \label{theorem:pure-nash-ratgenfun-encoding-concave}
Consider an integer programming game with DPLC payoffs given by the same input as in Theorem \ref{theorem:extended-main-result}. The set $N$ of pure Nash equilibria has a short rational generating function encoding which can be computed in polynomial time when the total dimension~$d$ and the sizes $|K_i|$ are fixed for all $i \in I$.
\end{corollary}
\begin{Proof}
By the previous theorem, we can encode the set of \emph{extended} pure Nash equilibria, $\hat N$, in polynomial time. Using the bijective map $\varphi$ given in the proof of  Lemma~\ref{lemma:bijection} we can use an appropriate monomial substitution in the rational generating function description of $\hat N$ to yield a short rational generating function encoding for $N$.
\end{Proof}

The true power of the previous lemma lies in the fact that having a short rational encoding of a set allows for efficient counting, optimizing and enumerating procedures as discussed in Section \ref{s:gen-fun}. Using these techniques we can answer numerous questions of interest on the existence, uniqueness and structure of equilibria.

\begin{corollary}\label{theorem:count-find-sample-Nash} Consider an integer programming game with DPLC payoffs given by the same input as in Theorem \ref{theorem:extended-main-result}. There is a polynomial time algorithm to compute the number of pure Nash equilibria, when the total dimension~$d$ and the sizes $|K_i|$ are fixed for all $i \in I$. In addition, under the same assumptions there is a polynomial time algorithm to find a sample pure strategy Nash equilibrium, when at least one exists.
\end{corollary}
\begin{Proof}
Given the short rational generating function encoding of $N$ we calculate $|N|$ in polynomial time by counting methods discussed near the beginning of Section \ref{s:gen-fun}. If an equilibrium exists, we can output one by running the polynomial-delay enumeration algorithm on $N$ described in the Enumeration Theorem and terminating just after one equilibrium has been generated. This can be done in polynomial time.
\end{Proof}

Note that the algorithm runs in time polynomial in the total number $\sum_{i \in I} |L_i|$ of ``concave" pieces in the payoff functions. This corollary can be used to answer the question of whether a unique pure Nash equilibrium exists -- simply check whether $|N|=1$. The following result is also immediate by the Enumeration Theorem.

\begin{corollary}\label{theorem:Nash-enumeration}
Consider an integer programming game with DPLC payoffs given by the same input as in Theorem \ref{theorem:extended-main-result}. There is a polynomial-space polynomial-delay algorithm to enumerate all the pure Nash equilibria of the game when the total dimension~$d$ and the sizes $|K_i|$ are fixed for all $i \in I$.
\end{corollary}

Reflecting on the content and context to the results of this section, we make the following useful remarks.

\begin{remark}
Considering the number of elements of the game we need to fix in Corollary~\ref{theorem:count-find-sample-Nash} -- fixed number of players, fixed dimension of the polytopes, fixed sizes of the $K_i$ -- one might ask if there is an alternate method to generating functions that might yield similar results. The key observation is that the action sets are described implicitly as lattice points in polytopes given by linear inequalities, and thus the number of actions for each player may be exponential in the input size. Thus, simple enumeration of all the action profiles in $S$ is a computationally intractable approach to the problem.
\end{remark}

\begin{remark}\label{remark:expressibility}
It was shown in Proposition \ref{prop:expressibility} that every normal form game can be expressed as an integer programming game with DPLC payoffs. Note, however, that the dimensions of the action spaces are equal to number of corresponding actions in the normal form game. Indeed, using the notation of the proof of Proposition~\ref{prop:expressibility}, we have $S_i \in \Z^{A_i}$. From a complexity point of view this representation is unsatisfactory. In Corollary~\ref{theorem:count-find-sample-Nash} we require the dimension of the action spaces to be fixed, and thus we can only handle a fixed number of actions in the underlying normal form game. Normal form games with a fixed number of players and fixed number of actions are computationally uninteresting.
\end{remark}

\begin{remark}\label{remark:example}
The Cournot game in Example \ref{example} fits the assumptions of Corollary~\ref{theorem:count-find-sample-Nash}. We assume that the number $n$ of players (manufacturers) is ``small", i.e., fixed, in order for each manufacturer to have appreciable market power. We also assume that the total number $d$ of products is small. The sets $K_i$ have cardinality $O(2^{d_i})$, which is fixed, and thus the decomposition of $S$ in Corollary~\ref{cor:partition} is comprised of a constant number of subsets. Since the algorithm in Corollary~\ref{theorem:count-find-sample-Nash} scales polynomially with the sizes of the sets $L_i$, we can afford a variable number of ``concave" pieces in the description of the payoff functions. These ``concave" pieces are used to represent general concave revenue functions and the convex parts of the cost functions, when restricted to integer points.
\end{remark}

\section{Related computations}\label{s:computations}

In addition to counting and enumerating pure Nash equilibria, generating function techniques can be used to derive efficient algorithms for related computations for integer programming games. Several of these are discussed in the following subsections.

First, note that the encoding of the set of Nash equilibria as a short rational generating function, being a compact representation, is useful for learning about the specific structure of a game's equilibria. For instance, a simple calculation suffices for deciding the existence of a pure Nash equilibrium where player $i$ plays a given action $\bar{\ve s}_i$. Indeed, simply find the short rational generating function encoding of
$$N^{(\bar{\ve s}_i)} = N \cap \{\ve s \in S: \ve s_i = \bar{\ve s}_i\}.$$
in polynomial time under the same assumptions in the previous section.

Now onto some more sophisticated calculations.

\subsection{Pareto optimality}\label{ss:pareto}

Consider the question of finding Pareto optimal pure Nash equilibria, if any exist, in an integer programming game with DPLC payoffs. To tackle this, we start by encoding the set of Pareto optimal action profiles of the game.
\begin{theorem}\label{theorem:pareto}
Consider an integer programming game with DPLC payoffs given by the same input as in Theorem \ref{theorem:extended-main-result}. The set $PO$ of Pareto optimal action profiles has a short rational generating function encoding, which can be computed in polynomial time when the total dimension~$d$ and the sizes $|K_i|$ are fixed for all $i \in I$.
\end{theorem}
\begin{Proof}
The proof is similar to those in the proof of Lemma~\ref{lemma:bijection}, Theorem~\ref{theorem:extended-main-result} and Corollary~\ref{theorem:pure-nash-ratgenfun-encoding-concave}:
\begin{enumerate}[(i)]
\item Define $\widehat{PO}$ as the set of Pareto optimal points in the extended game and find a generating function encoding for $\widehat{PO}$.\label{aa}
\item Derive a bijection between $PO$ and $\widehat{PO}$.\label{bb}
\item Use the generating function of $\widehat {PO}$ and the bijection to obtain the generating function of $PO$.\label{cc}
\end{enumerate}
For part \eqref{aa} consider the following decomposition of $\widehat{PO}$:
\begin{eqnarray*}
\widehat{PO} & = & \{\hat {\ve s} \in \hat S: \nexists~\hat{\ve s}^\prime \in \hat S \;	\text{ such that }	
											(\hat u_j (\hat {\ve s}^\prime)  \ge  \hat u_j (\hat {\ve s}) \text{ for all } j \in I) \text{ and }
											(\hat u_i (\hat {\ve s}^\prime)  >  \hat u_i (\hat {\ve s}) \text{ for some } i \in I) \} \\
                & = & \hat S \setminus \bigcup_{i=1}^n PD_i
\end{eqnarray*}
where
\begin{equation*}
PD_i = \bigcup_{\ve k, \ve k^\prime \in \ve K} \proj_{\hat {\ve s}}\left\{(\hat {\ve s}, \hat {\ve s}^\prime):
\hat {\ve s} \in  \hat S_{\ve k},~~
\hat {\ve s}^\prime \in   \hat S_{\ve k'},~~
\hat u_j(\hat{\ve s}^\prime) \ge  \hat u_j(\hat {\ve s}) ~~ \text{ for all } j \not = i, ~~
\hat u_i(\hat {\ve s}^\prime)  \ge   \hat u_i(\hat {\ve s})+1
\right\}
\end{equation*}
is the set of Pareto dominated points due to a better alternative for player $i$. By analogous arguments an in the proof of Theorem~\ref{theorem:extended-main-result} we can encode $\widehat{PO}$ as a short rational generating function in polynomial time.

As for \eqref{bb} the argument is nearly identical to that in the proof of Lemma~\ref{lemma:bijection} and is therefore omitted. Finally, \eqref{cc} uses the same idea as found in the proof Corollary~\ref{theorem:pure-nash-ratgenfun-encoding-concave}.
\end{Proof}

Using the Boolean Operations Theorem we obtain a short rational generating function encoding of the set $N \cap PO$ of all Pareto optimal pure Nash equilibria of the original game:

\begin{corollary}\label{cor:pareto-nash}
Consider an integer programming game with DPLC payoffs given by the same input as in Theorem \ref{theorem:extended-main-result}. The set of Pareto optimal pure Nash equilibria has a short rational generating function encoding, which can be computed in polynomial time when the total dimension $d$ and the sizes $|K_i|$ are fixed for all $i \in I$.
\end{corollary}

As in Section \ref{s:pure-nash} we can use this generating function encoding to count and enumerate Pareto optimal equilibria.

\subsection{Pure prices of anarchy and stability}\label{ss:anarchy}

The pure price of anarchy of a game measures the negative effects of competition on social welfare. The \emph{social welfare} of an action profile is the corresponding total payoff of all players. The \emph{pure price of anarchy} is the ratio of the maximum social welfare where the agents act together to maximize their total payoff to the \emph{worst} social welfare that arises from a pure Nash equilibrium. The \emph{pure price of stability} is the ratio of maximum social welfare to the \emph{best} social welfare that arises from a pure Nash equilibrium. The pure price of anarchy has been studied in various network games, see \cite{Dunkel2008} and the references therein for recent results on weighted congestion games. Using rational generating function techniques we can calculate the pure price of anarchy and stability efficiently.

\begin{theorem}\label{theorem:price}
Consider an integer programming game with DPLC payoffs given by the same input as in Theorem \ref{theorem:extended-main-result}. There exist algorithms to compute the pure price of anarchy and the pure price of stability that run in polynomial time when the total dimension~$d$ and the sizes $|K_i|$ are fixed for all $i \in I$.
\end{theorem}
\begin{Proof}
Let $w^*$ denote the maximum social welfare attainable under cooperation of the players; that is,
\begin{equation}\label{11} w^* = \max \left\{ \sum_{i=1}^n u_i(\ve s): \ve s \in S\right\}.\end{equation}

Again we work with the extended game. The first step in calculating $w^*$ is to note that
\begin{equation}\label{22}
w^* =  \max \left\{\sum_{i=1}^n y_i : (\ve s; \ve y) \in \hat S\right\}
\end{equation}
The equivalence of \eqref{11} and \eqref{22} is verified by first noting that for all $(\ve s; \ve y) \in \hat S$, $y_i \le u_i(\ve s) = \sum_i \hat u_i(\ve s;u(\ve s))$.
Therefore, if $(\ve s^*, \ve y^*)$ is an optimal solution to the right-hand side of \eqref{22} then we must have $y_i^* = u_i(\ve s)$ for all $i \in N$. This implies $\ve s$ is an optimal solution the right-hand side of \eqref{11}.

To find $w^*$ we optimize the linear function $\sum_i y_i$ over $\hat S$. Every $(\ve s, \ve y) \in \hat S$ satisfies $\ve s \in [-B,B]^d$ and
 \[y_i \le u_i(\ve s) \le B\left(\max_{k\in K_i}(||\ve \alpha_{ik}||_1 + |\beta_{ik}|) + \max_{l\in L_i}(||\ve \gamma_{il}||_1 + |\delta_{il}|)\right).\]
 That is, $(\ve s; \ve y) \in [-M,M]^{d+n}$ for some polynomially sized integer $M$.
 Since in Section \ref{s:pure-nash} we found a short rational generating function encoding of $\hat S$, we apply the Linear Optimization Theorem to calculate $w^*$ in polynomial time.

Let $\tilde{w}$ denote the worst social welfare attained by a pure Nash equilibrium; that is,
$$\tilde{w} = \min \left\{ \sum_{i=1}^n u_i(\ve s): \ve s \in N\right\}.$$
To calculate $\tilde{w}$ we note
$$\tilde{w} = \min \left\{\sum_{i=1}^n y_i : (\ve s;\ve y) \in \hat N\right\},$$
which follows from Lemma~\ref{lemma:bijection}(i).
Using the short generating function encoding of $\hat N \subseteq \hat{S}$ found in Section \ref{s:pure-nash} we again apply the Linear Optimization Theorem to calculate $\tilde{w}$ in polynomial time. Thus, we obtain the price of anarchy $w^*/\tilde{w}$ in polynomial time.

The method for calculating the pure price of stability is similar.
\end{Proof}

\subsection{Pure Threat Point}\label{ss:threat}
The \emph{pure minmax value to player $i$} in a game is defined as:
\begin{equation}\label{eq:minmax-value}
\min_{\ve s_{-i} \in S_{-i}} ~ \max_{\ve s_i \in S_i} ~u_i(\ve s_i, \ve s_{-i}).
\end{equation}
Although mixed strategies are usually considered in calculating the (mixed) minmax values, here we restrict attention to pure strategies.  The vector of \emph{mixed} minmax values is known as the (mixed) \emph{threat point}, which has drawn recent attention in the study of repeated games and explorations of computational implications of the Folk Theorem (see \cite{Borgs2007}). Analogously, we define the \emph{pure threat point} as the vector of pure minmax values. %The pure threat point plays the same role in repeated games restricted to pure strategies as the (mixed) threat point when mixed strategies are allowed.

It was recently shown that, in various restrictive settings, the problem of calculating the (mixed) threat point is NP-hard. For instance, it can be shown that computing the (mixed) threat point of a three player game with binary payoffs (\{0,1\}) is NP-hard to approximate (see \cite{Borgs2007} Theorem 1 for a precise statement and proof.) Despite this negative result we show that pure threat points can be computed efficiently in our setting.

\begin{theorem}\label{theorem:threat}
Consider an integer programming game with DPLC payoffs given by the same input as in Theorem \ref{theorem:extended-main-result}. There exists a polynomial time algorithm to compute the \emph{pure} threat point when the total dimension~$d$ and the sizes $|K_i|$ are fixed for all $i \in I$.
\end{theorem}
\begin{Proof}
We begin by demonstrating how to calculate the minmax value for player $i$ in polynomial time for each player $i$. Observe that an optimal value to the following bilevel optimization problem is the pure minmax value of player $i$:
\begin{equation}\label{eq:bilevel-minmax}
\min_{\ve s_i,\ve s_{-i}}  \left\{u_i(\ve s_i,\ve s_{-i}) :  \ve s_{-i} \in S_{-i}, ~\ve s_i \in \arg \max_{\ve s'_i \in S_i} u_i(\ve s'_i,\ve s_{-i})\right\}.
\end{equation}
This bilevel optimization problem (see \cite{Colson2007}) has essentially two players: a lower level player, or follower, who is player $i$, and an upper level player, or leader, who represents all the other players cooperating to ``punish" $i$. Let
$$G_i = \big\{\ve s \in S: \ve s_i \in \arg \max_{{\ve s}_i \in S_i} \{u_i(\ve s_i,\ve s_{-i})\}\big\}$$
denote the set of bilevel feasible solutions to (\ref{eq:bilevel-minmax}). Note that \eqref{eq:bilevel-minmax} is equivalent to $\min \{u_i(\ve s): \ve s \in G_i\}$.

As before we turn our attention to the extended game. We define the analogous set $\hat G_i$:
\begin{equation*}
\hat G_i   =  \left\{\hat {\ve s} \in \hat S: \hat{\ve s_i} \in \arg \max_{\hat{\ve s}_i^\prime} \{\hat u_i(\hat{\ve s}_{-i}, {\hat{\ve s}_i}):   (\hat{\ve s}_{-i}, \hat{\ve s}_i^\prime) \in \hat S\}\right\}. \\
\end{equation*}
Observe that if $\hat {\ve s} = (\ve s, \ve y) \in \hat G_i$ then $y_i = u_i(\ve s)$.
The set $\hat G_i$ can be expressed as $\hat G_i = \hat S \setminus D_i$,
where $D_i$ is defined as in \eqref{eq:deviation-set}. This follows since the optimization problem facing player $i$ is the same problem as when determining extended Nash equilibria is a single player game.
Thus by a direct application of Theorem~\ref{theorem:extended-main-result} we can encode $\hat{G}_i$ as a short rational generating function. Note that $\hat G_i \subseteq \hat S \subseteq [-M,M]^{d+n}$ where $M$ is as defined in the proof of Theorem~\ref{theorem:price}. By applying the Linear Optimization Theorem find the optimal value of $\min{y_i: (\ve s, \ve y) \in \hat {G}_i = \min \{u_i(\ve s}): \ve s \in G_i\}$, in polynomial time under the stated assumptions. The pure threat point can thus be calculated in polynomial time by finding the minmax value for each player.
\end{Proof}

\section{Stackelberg--Nash equilibria}\label{s:stack-nash}

We now turn to applying these techniques to a sequential setting. In a \emph{Stackelberg--Nash game}, Player~$0$ (the leader) chooses an action, described by a vector $\ve s_0 \in S_0$. The remaining players~$i\in I =\{1,\ldots,n\} $ (the followers)
then simultaneously choose their actions $\ve s_i \in S_i(\ve s_0)$. Each player $i \in I_0 = \{0,1,\ldots,n\}$ then collects a payoff $u_i(\ve s_0, \ve s)$ where $\ve s = (\ve s_1,\ldots, \ve s_n)$.

We assume $S_0 = P_0 \cap \Z^{d_0}$ where $P_0$ is a rational polytope. For each $\ve s_0 \in S_0$ the followers play an integer programming game with DPLC payoffs. The action of each follower $i\in I$ is described by a vector~$\ve s_i \in \Z^{d_i}$
from the set $S_i(\ve s_0) = P_i(\ve s_0) \cap \Z^{d_i}$. We assume $P_i(\ve s_0) $ is the rational polytope
$$P_i(\ve s_0) = \{\ve x \in \R^{d_i}: M_i\ve x \le \pi_i(\ve s_0)\} ~ \text{for \ } ~ i \in I$$
where $\pi_i(\ve s_0)$ is an integer valued affine function. Let $d = d_1 + \dots + d_n$ and $d^+ = d_0 + d$.

Regarding payoffs, we assume each follower has a DPLC payoff $u_i(\ve s)$, independent of the leader's choice $\ve s_0$ and given by
\begin{equation}\label{eq:DPLC-followers}
u_i(\ve s) = \max_{k\in K_i} f_{ik}(\ve s) - \max_{l \in L_i} g_{il}(\ve s).
\end{equation}
The leader's payoffs are defined as the DPLC function
\begin{equation}\label{eq:DPLC-leader}
u_0(\ve s_0,\ve s) = \max_{k\in K_0} f_{0k}(\ve s_0, \ve s) - \max_{l \in L_0} g_{0l}(\ve s_0, \ve s).
\end{equation}
We assume all $K_i$ and $L_i$ are finite index sets and all $f_{ik}$ and $g_{il}$ are integer valued affine functions.

Observe that given $\ve s_0 \in S_0$ we have a setup identical to that of Section \ref{s:pure-nash}, where the set of action profiles for the followers is $S(\ve s_0) = \prod_{i=1}^n S_i(\ve s_0) \subseteq \Z^d$.

We are interested in computing an optimal action for the leader while guaranteeing there exists a pure Nash equilibrium between the
followers; see Remark \ref{remark:focus-on-pure} below for justification of our interest in pure Nash equilibria.
Let $N(\ve s_0)$ denote the set of pure Nash equilibria between the
followers when the leader has chosen action~$\ve s_0\in S_0$.
As in Section \ref{s:pure-nash}, a pure Nash equilibrium in $N(\ve s_0)$
is an action profile $\ve s \in S(\ve s_0)$ such that for every $i \in I$ there does not exist a deviation $s_i^\prime \in S_i(\ve s_0)$ such that $u_i(\ve s_{-i},\ve s_i^\prime) > u_i(\ve s)$.

The leader faces the following optimization problem:
\begin{equation}\label{eq:Stackelberg--Nash}
\max_{\ve s_0, \ve s}\{u_0(\ve s_0, \ve s) : s_0 \in S_0 \text{ and } s \in N(\ve s_0)\}.
\end{equation}
Let $N^+$ denote the set of all \emph{Stackelberg--Nash equilibria}, i.e., optimal
solutions~$(\ve s_0; \ve s)$ to the optimization
problem~\eqref{eq:Stackelberg--Nash}.

\begin{remark}\label{remark:optimistic}
Note that this formulation implicitly assumes that the leader, after choosing $\ve s_0$, can choose a pure Nash equilibrium $\ve s \in N(\ve s_0)$ in order to maximize her payoff. This is a generalization to the case of competing followers of the ``optimistic assumption''  common in the multilevel optimization literature. The simplest illustration of the assumption is in the bilevel setting where the leader has the ability to choose among alternate optima to the follower's problem (see \cite{Colson2007}). Here we assume more generally that the leader can choose among the alternate pure Nash equilibria between the followers.
\end{remark}

\begin{remark}\label{remark:focus-on-pure}
The focus solely on pure strategies may need some motivation. Some choice of $\ve s_0 \in S_0$ may give rise to no pure Nash equilibria in the followers' game, leaving only mixed Nash equilibria. We assume that the leader will avoid such an $\ve s_0$, even if it gave rise to higher \emph{expected} payoffs. Consider this as an extreme form of risk aversion, where any equilibrium in pure strategies in preferred by the leader so as to avoid any uncertainty in payoffs.  By similar reasoning, we also assume the leader will not be interested in the mixed equilibria when pure equilibria exist. Extending the optimistic assumption discussed in Remark~\ref{remark:optimistic} we assume the leader can compel the followers to reach a pure equilibrium whenever it exists.
\end{remark}

\begin{theorem}\label{theorem:stack-nash}
 Consider a Stackelberg--Nash game with DPLC payoffs defined by the following
 input, given in binary encoding:
 \begin{inputlist}
\item the number $n$ of followers, and a bound $B \in \N$;
\item the dimension $d_0$ and an inequality description $(M_0,\ve b_0)$ of a rational polytope $P_0 = \{\ve x \in \R^{d_0}: M_0\ve x \le \ve b_0\}\subseteq [-B,B]^{d_0}$ defining the leader's feasible set $S_0 = P_0 \cap \Z^{d_0}$;
\item for each $i \in I = \{1,\dots,n\}$, the dimension $d_i$, number $m_i$ of constraints, integer $m_i\times d_i$ matrix $M_i$, integer $d_0 \times m_i$ matrix $\Phi_{i}$ and integer vector $\psi_{i} \in \Z^{m_i}$ defining the affine function $\pi_{i}:S_0 \rightarrow \Z^{m_i}$ by $\pi_{i}(\ve s_0) = \Phi_i \ve s_0 + \psi_{i}$, and defining the follower $i$'s parameterized polytope $P_i(\ve s_0) = \{\ve x \in \R^{d_i}: M_i\ve x \le \pi_i(\ve s_0)\}$;
\item for each $i \in I$, nonnegative integers $|K_i|$ and $|L_i|$, and for all integers $k$, $l$ such that $1\le k \le |K_i|$ and $1\le j \le |L_i|$, integer vectors $\ve \alpha_{ik} \in \Z^{d}$, $\ve \gamma_{il} \in \Z^{d}$ (where $d = d_1 + \cdots + d_n$) and integers $\beta_{ik}$, $\delta_{ik}$   defining the affine functions $f_{ik}:\Z^{d} \rightarrow \Z$ and $g_{il}:\Z^{d}\rightarrow \Z$ by $f_{ik}(\ve s) = \ve \alpha_{ik}\cdot\ve s + \beta_{ik}$ and $g_{il}(\ve s) = \ve \gamma_{il}\cdot\ve s + \delta_{il}$ for all $\ve s\in \Z^{d}$;
\item nonnegative integers $|K_0|$ and $|L_0|$, and for all integers $k$, $l$ such that $1\le k \le |K_0|$ and $1\le j \le |L_0|$, integer vectors $\ve \alpha_{0k} \in \Z^{d^+}$, $\ve \gamma_{0l} \in \Z^{d^+}$ (where $d^+ = d_0 + d$) and integers $\beta_{0k}$, $\delta_{0k}$  defining the affine functions $f_{0k}:\Z^{d^+} \rightarrow \Z$ and $g_{0l}:\Z^{d^+}\rightarrow \Z$ by $f_{0k}(\ve s_0, \ve s) = \ve \alpha_{ik}\cdot(\ve s_0,\ve s) + \beta_{0k}$ and $g_{0l}(\ve s_0,\ve s) = \ve \gamma_{0l}\cdot(\ve s_0,\ve s) + \delta_{0l}$ for all $(\ve s_0,\ve s) \in \Z^{d^+}$.
\end{inputlist}
Then there exists a polynomial-time
algorithm to compute the leader's optimum payoff and a short rational generating
function encoding of the set~$N^+$ of all Stackelberg--Nash equilibria when the total dimension $d^+$ and the sizes $|K_0|,|K_1|, \dots,|K_n|$ are fixed. % and $|S_0|$ is polynomial in the input size.
\end{theorem}
\begin{Proof}
We mimic the development leading up to Theorem~\ref{theorem:extended-main-result} in Section~\ref{s:pure-nash} by defining an extended game with extended strategy profiles $(\ve s_0, \ve s, \ve y)$ where $y_i \le u_i(\ve s)$ for all $i \in I$. As before denote $\hat{\ve s} = (\ve s, \ve y)$. Let
\begin{displaymath}
S_{ik} = \left\{(\ve s_0, \ve s) : \ve s_0 \in S_0,~\ve s \in S(s_0), ~ f_{ik}(\ve s) \ge f_{ij}(\ve s) \text{ for } ~j > k,~	f_{ik}(\ve s) > f_{ij}(\ve s) \text{ for } ~j <k
\right\}
\end{displaymath}
and thus construct the disjoint union
\begin{displaymath}
\hat S = \biguplus_{\ve k \in \ve K} \hat S_{\ve k}
\end{displaymath}
where
\begin{displaymath}
\hat S_{\ve k} = \{(\ve s_0, \ve s, \ve y): \ve s_0 \in S_0,~ \ve s \in  S_{\ve k}  \text{ and } 0 \le y_i \le f_{ik}(\ve s) - g_{il}(\ve s) \text{ for all } l \in L_i \text{ and } i \in I \}.
\end{displaymath}
denoting $S_{\ve k} = S_{1k_1}\cap \cdots \cap S_{nk_n}$. Note that $\hat S_{\ve k}$ is a lattice point set in a polytope, and thus we can encode $\hat S$ by a short rational generating function.

Let $\hat{N}$ denote the set of extended action profiles $(s_0;\hat{s})$ such that $s_0\in S_0$ is any feasible leader action and $\hat{s}$ is a pure Nash equilibrium in the followers extended game when the leader has chosen $s_0$. Now express $\hat N$ as
\begin{displaymath}
\hat N = \hat S \setminus \bigcup_{i=1}^n D_i
\end{displaymath}
where
\begin{eqnarray*}
D_i = \biguplus_{\ve k\in \ve K}~\bigcup_{k_i^\prime \in K_i} \proj_{\ve s_0, \hat {\ve s}}\left\{(\ve s_0, \hat {\ve s}, \hat {\ve s}_i^\prime): \begin{array}{lll}
(\ve s_0, \hat{\ve s}) \in \hat S_{\ve k}, \\
(\ve s_0, \hat{\ve s}_{-i},\hat{\ve s}'_i  \in  \hat S_{(\ve k_{-i},k_i^\prime)}, \\
      ~ \hat u_i(\hat {\ve s}_{-i}, \hat {\ve s}_i^\prime) \ge \hat u_i(\hat {\ve s}) + 1
\end{array} \right\}.
\end{eqnarray*}
Since all sets to be projected are lattice point sets inside polytopes, we can apply reasoning as in the proof of Theorem~\ref{theorem:extended-main-result} and encode $\hat N$ by a short rational generating functions under the stated assumptions. We establish a bijection between $\hat N$ and the set $N = \{(\ve s_0, \ve s) : \ve s_0 \in S_0,~\ve s \in N(\ve s_0)\}$ of feasible solutions to \eqref{eq:Stackelberg--Nash}. We claim the function $\varphi : N \longrightarrow \hat{N}$ defined by $(\ve s_0, \ve s) \longmapsto (\ve s_0, \ve s; u(\ve s))$ is a well-defined bijection. The details are similar to that in the proof of Lemma~\ref{lemma:bijection} and are thus omitted. Via this bijection we derive a short rational generating function encoding of $N$ by monomial substitution.

Now, let $N^{(v)}$ denote the set of feasible solutions to \eqref{eq:Stackelberg--Nash} that guarantee the leader a payoff of at least~$v$:
\begin{eqnarray*}
N^{(v)} & =  & N \cap \left(\{(\ve s_0,\ve s): u_0(\ve s_0, \ve s) \ge v\}\right) \\
              & = & N \cap \left(\bigcup_{k \in K_0} \{(\ve s_0, \ve s):  f_{0k}(\ve s_0, \ve s) - g_{0l}(\ve s_0, \ve s) \ge v ~~\forall l \in L_0\} \right).
\end{eqnarray*}
Note this is an intersection and union of polytopal lattice point sets. Since the size of $K_0$ is fixed we can apply the Boolean Operations Theorem to encode $N^{(v)}$ by a short rational generating function. Now, using binary search for $v$ between $-B^+$ and $B^+$ where
\[B^+ = B\left(\max_{k\in K_0}(||\ve \alpha_{0k}||_1 + |\beta_{0k}|) + \max_{l\in L_0}(||\ve \gamma_{0l}||_1 + |\delta_{0l}|)\right),\]
and the counting algorithm to test for non-emptiness of $N^{(v)}$, we can find an optimal payoff $v^+$ to the leader. The set $N^+$ is therefore equal to $N^{(v^+)}$.
\end{Proof}

As in the previous section, a short rational generating function encoding of the set~$N^+$ of Stackelberg--Nash equilibria leads to results analogous to Corollary~\ref{theorem:count-find-sample-Nash} and Corollary~\ref{theorem:Nash-enumeration}. Thus, we can derive efficient procedures to decide on the existence of and to enumerate Stackelberg--Nash equilibria.

\section{Conclusions and directions for further research}

In this paper we introduced classes of games and proposed algorithms for studying their pure strategy Nash equilibria using rational generating functions. The simplicity by which they can be used to compute important information on the structure of these games, demonstrates the power of generating functions as an analytical and computational tool for game theory.

There is considerable scope to explore applications of integer programming games to other situations. In addition, the use of generating function techniques is a novel approach to game theory and can be applied to various types of games; e.g., threat point computations for repeated integer programming games; games with other payoff functions; and computations related to other solution concepts.

\section*{Acknowledgements}

The research of the last two authors was supported in part by a Discovery grant from the Natural Sciences and Engineering Research Council (NSERC) of Canada to the third author. The second author is also supported in part by the Shelby L. Brumelle Memorial Scholarship.

\bibliographystyle{ormsv080}
\bibliography{games_references}

\end{document}